\documentclass[prl,twocolumn,superscriptaddress,showpacs,times,10pt]{revtex4-1}
\usepackage{graphicx,amsmath,amssymb,bm}
\usepackage{color}
\usepackage{array}
\usepackage{rotating,pifont}
\usepackage{simplewick}

\DeclareMathSymbol{\NS}{\mathord}{AMSb}{"4E}

\newcommand{\ket}[1]{\ensuremath{\,|{#1}\rangle}}

\newcommand{\op}[1]{\ensuremath{#1}}
\newcommand{\adj}[1]{\ensuremath{{{#1}}^{\dag}}}

\renewcommand{\vec}[1]{\ensuremath{\bm{#1}}}

\newcommand{\aaO}{\ensuremath{\adj{\op{a}}}}

\newcommand{\TO}{\ensuremath{\op{T}}}

\newcommand{\pOV}{\ensuremath{\vec{\op{p}}}}

\newcommand{\Tint}{\ensuremath{\TO_\text{int}}}

\newcommand{\hw}{\ensuremath{\hbar\omega}}
\newcommand{\lambdaSRG}{\ensuremath{\lambda_{\text{SRG}}}}

\newcommand{\eMax}{\ensuremath{e_{\text{max}}}}

\newcommand{\EMax}{\ensuremath{E_{3\text{max}}}}

\newcommand{\nuc}[2]{\ensuremath{{}^{#2}\mathrm{#1}}}

\newcommand{\fm}{\ensuremath{\,\text{fm}}}
\newcommand{\fmi}{\ensuremath{\,\text{fm}^{-1}}}

\newcommand{\keV}{\ensuremath{\,\text{keV}}}
\newcommand{\MeV}{\ensuremath{\,\text{MeV}}}

\newcommand{\nord}[1]{\ensuremath{:\!#1\!:}}

\newcommand{\df}{\ensuremath{d_{5/2}}}
\newcommand{\dt}{\ensuremath{d_{3/2}}}
\newcommand{\so}{\ensuremath{s_{1/2}}}
\newcommand{\fs}{\ensuremath{f_{7/2}}}
\newcommand{\pt}{\ensuremath{p_{3/2}}}

\newcommand{\tpi}{\ensuremath{2^+_1}}

\definecolor{FGViolet}{rgb}{0.61,0.32,0.61}
\definecolor{FGDarkBlue}{rgb}{0,0,0.6}
\definecolor{FGBlue}{rgb}{0,0,0.8}
\definecolor{FGLightBlue}{rgb}{0.2, 0.6, 0.8}
\definecolor{FGGreen}{rgb}{0.2,0.7,0.2}
\definecolor{FGLightGreen}{rgb}{0.4,1,0.4}
\definecolor{FGYellow}{rgb}{1,0.95,0}
\definecolor{FGOrange}{rgb}{0.95,0.5,0.1}
\definecolor{FGRed}{rgb}{0.8,0,0}
\definecolor{FGWhite}{rgb}{1,1,1}
\definecolor{FGLightGray}{rgb}{0.8,0.8,0.8}
\definecolor{FGGray}{rgb}{0.5,0.5,0.5}
\definecolor{FGDarkGray}{rgb}{0.3,0.3,0.3}
\definecolor{FGBlack}{rgb}{0,0,0}

\begin{document}

\title{Nonperturbative shell-model interactions from the\\ in-medium similarity renormalization 
group}

\author{S.\ K.\ Bogner}
\email[E-mail:~]{bogner@nscl.msu.edu}
\affiliation{National Superconducting Cyclotron Laboratory
and Department of Physics and Astronomy, Michigan State University,
East Lansing, MI 48844, USA}

\author{H.\ Hergert}
\email[E-mail:~]{hergert.3@osu.edu}
\affiliation{Department of Physics, The Ohio State University, 
Columbus, OH 43210, USA}

\author{J.\ D.\ Holt}
\email[E-mail:~]{jason.holt@physik.tu-darmstadt.de}
\affiliation{ExtreMe Matter Institute EMMI, GSI Helmholtzzentrum f\"ur
Schwerionenforschung GmbH, 64291 Darmstadt, Germany}
\affiliation{Institut f\"ur Kernphysik, Technische Universit\"at
Darmstadt, 64289 Darmstadt, Germany}
\affiliation{National Superconducting Cyclotron Laboratory
and Department of Physics and Astronomy, Michigan State University,
East Lansing, MI 48844, USA}

\author{A.\ Schwenk}
\email[E-mail:~]{schwenk@physik.tu-darmstadt.de}
\affiliation{ExtreMe Matter Institute EMMI, GSI Helmholtzzentrum f\"ur
Schwerionenforschung GmbH, 64291 Darmstadt, Germany}
\affiliation{Institut f\"ur Kernphysik, Technische Universit\"at
Darmstadt, 64289 Darmstadt, Germany}

\author{S.\ Binder}
\affiliation{Institut f\"ur Kernphysik, Technische Universit\"at
Darmstadt, 64289 Darmstadt, Germany}

\author{A.\ Calci}
\affiliation{Institut f\"ur Kernphysik, Technische Universit\"at
Darmstadt, 64289 Darmstadt, Germany}

\author{J.\ Langhammer}
\affiliation{Institut f\"ur Kernphysik, Technische Universit\"at
Darmstadt, 64289 Darmstadt, Germany}

\author{R.\ Roth}
\affiliation{Institut f\"ur Kernphysik, Technische Universit\"at
Darmstadt, 64289 Darmstadt, Germany}

\begin{abstract}
We present the first \emph{ab initio} construction of valence-space Hamiltonians for 
medium-mass nuclei based on chiral two- and three-nucleon interactions using the 
in-medium similarity renormalization group.  When applied to the oxygen isotopes, we find
experimental ground-state energies are well reproduced, including the flat trend beyond the drip 
line at $^{24}$O.  Similarly, natural-parity spectra in $^{21, 22, 23, 24}$O are in agreement with
experiment, and we present predictions for excited states in
$^{25,26}$O.  The results exhibit a weak dependence on the harmonic-oscillator (HO) basis 
parameter and reproduce spectroscopy within the standard $sd$ valence space.  
\end{abstract}


\maketitle

With the next generation of rare-isotope beam facilities, the quest to discover and understand 
the properties of exotic nuclei from first principles is a fundamental challenge for nuclear theory. 
This challenge is complicated in part because the proper inclusion of three-nucleon (3N) forces 
plays a decisive role in determining the structure of medium-mass nuclei 
\cite{Otsuka:2009,Holt:2010}.  While \emph{ab initio} many-body methods based on nuclear 
forces from chiral effective field theory (EFT) 
\cite{chiralEFT,Machleidt:2011bh,Hammer:2013nx} have now reached the medium-mass region 
and beyond \cite{Hagen:2010uq,Hagen:2007zc,Hagen:2012a,Hagen:2012b,Roth:2012qf,Soma:2013ys,Binder:2013zr,Hergert:2013mi,Hergert:2013ij,Cipollone,Jansen:2013zr,Soma,Binder:2013uq,Hagen:2013kx,Lahde}, 
restrictions in the nuclei and observables accessible to these methods have limited their
application primarily to ground-state properties in semi-magic isotopic chains.  

For open-shell systems, rather than solving the full $A$-body problem, it is profitable to follow 
the shell-model paradigm by constructing and diagonalizing an effective Hamiltonian in which 
the active degrees of freedom are $A_v$ valence nucleons confined to a few orbitals near the 
Fermi level.  Both phenomenological and microscopic implementations of the shell model have 
been used with success to understand and predict the evolution of shell structure, properties of 
ground and excited states, and electroweak transitions 
\cite{Brown:2001,SMRMP,Otsuka:2013vn}. 

Recent microscopic shell-model studies have revealed the impact of 3N forces in predicting 
ground- and excited-state properties in neutron- and proton-rich nuclei 
\cite{Otsuka:2009,Holt:2010,Holt:2011fj,Holt:2012,Holt:2013,Gallant,Wienholtz}. 
Despite the novel insights gained from these studies, they make approximations which are 
difficult to benchmark. The microscopic derivation of the effective valence-space Hamiltonian 
relies on many-body perturbation theory (MBPT) \cite{MBPT}, where order-by-order 
convergence is unclear. Even with efforts to calculate particular classes of diagrams 
nonperturbatively \cite{Holt:2005}, results are sensitive to the HO frequency $\hw$ (due to the 
core), and the choice of valence space \cite{Holt:2010,Holt:2011fj,Holt:2012}.  A 
nonperturbative method to address these issues was developed in 
\cite{Lisetskiy:2008,Lisetskiy:2009}, which generates valence-space interactions and operators 
by projecting their full no-core shell model (NCSM) counterparts into a given valence space. 

To overcome these limitations in heavier systems, the in-medium similarity renormalization 
group (IM-SRG), originally developed for \emph{ab initio} calculations of 
ground states in closed-shell systems \cite{Tsukiyama:2011}, can be extended to derive 
effective valence-space Hamiltonians and operators nonperturbatively. 
Calculations without initial 3N forces 
\cite{Tsukiyama:2012} indicated that an \emph{ab initio} description of ground and 
excited states for open-shell nuclei may be possible with this approach. In this Letter, we apply 
the IM-SRG starting from chiral nucleon-nucleon (NN) and 3N forces to the more challenging 
and physically interesting problem of the oxygen isotopes.

The IM-SRG is a continuous unitary transformation $U(s)$, parameterized by a flow parameter 
$s$, that drives the Hamiltonian to a band- or block-diagonal form \cite{SRG}. This is 
accomplished by solving the flow equation
\begin{equation}
\label{eq:srg}
\frac{dH(s)}{ds} =[\eta(s),H(s)]\,, 
\end{equation}
where $\eta(s) \equiv [dU(s)/ds]\,U^{\dagger}(s)$ is the anti-Hermitian generator of the 
transformation. With a suitable choice of $\eta(s)$, the off-diagonal part of the Hamiltonian, 
$H^{\rm{od}}(s)$, is driven to zero as $s$ approaches $\infty$. The ``in-medium'' label indicates
that we control the proliferation of induced many-body operators by normal 
ordering the Hamiltonian with respect to a finite-density reference state for each 
system of interest. We truncate Eq.~\eqref{eq:srg} to normal-ordered two-body operators, which 
we refer to as the IM-SRG(2) approximation. Initial results for $\nuc{Li}{6}$ agreed well with 
NCSM \cite{Tsukiyama:2012}, and a quantitative comparison with the importance-truncated NCSM 
\cite{Roth:2011kx} is underway.

The utility of the IM-SRG lies in the freedom to tailor the definition of $H^{\rm{od}}$ to a specific 
problem.  For instance, to construct a shell-model Hamiltonian for a nucleus comprised of 
$A_v$ valence nucleons outside a closed core, we define a Hartree-Fock (HF) reference state
$\ket{\Phi}$ for the core with $A_c$ particles, and split the single-particle basis into hole ($h$),
valence ($v$), and non-valence ($q$) particle states.  Treating all $A$ nucleons as active,
we eliminate matrix elements which couple $\ket{\Phi}$ to 
excitations, just as in IM-SRG ground-state calculations 
\cite{Tsukiyama:2011,Hergert:2013mi,Hergert:2013ij}. In addition, we decouple states with 
$A_v$ particles in the valence space, 
\mbox{$:\!\aaO_{v_1}\ldots\aaO_{v_{A_v}}\!\!:\!\!\ket{\Phi}$}, from states containing non-valence 
states. 

Normal-ordering the Hamiltonian with respect to $\ket{\Phi}$ and working in the IM-SRG(2) 
truncation
\begin{equation}
H(s) = E_0 + \sum_{ij} f_{ij} \, \nord{a_i^\dagger a_j}
+ \frac{1}{4} \sum_{ijkl} \Gamma_{ijkl} \, \nord{a_i^\dagger a_j^\dagger
a_l a_k} \,,
\end{equation}
we define \cite{Tsukiyama:2012}
\begin{equation}\label{eq:def_HodA}
  \left\{H^{\rm{od}}\right\} = \left\{f_{ph}, f_{pp'},f_{hh'}, \Gamma_{pp'hh'}, \Gamma_{pp'vh}\,,\Gamma_{pqvv'}\right\} + \text{H.c.}
\end{equation}
and use the White generator defined in 
Refs.~\cite{Hergert:2013mi, Tsukiyama:2011,Tsukiyama:2012}.  With this choice of generator, 
$H^{\rm{od}}(\infty)\rightarrow 0$, and the shell-model Hamiltonian is obtained by taking all 
valence-space matrix elements. 

\begin{figure*}[t]
\includegraphics[scale=.66,clip=]{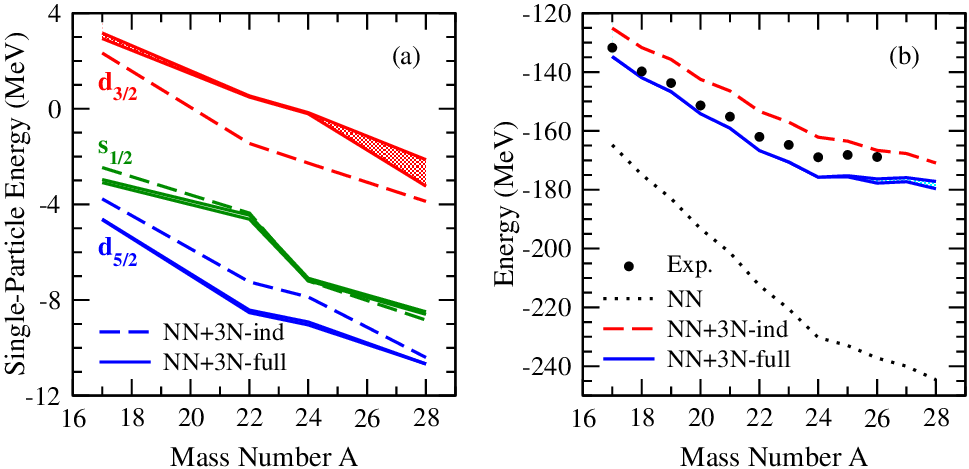}
\caption{Single-particle energy evolution (a) and ground-state energies (b) for $A$-dependent 
Hamiltonian with $\lambdaSRG = 1.88 \fmi$. The range of NN+3N-full ($\Lambda_{\rm 3N}=400\MeV$) results for $\hw=20,24\MeV$ is given by the shaded bands.
\label{gs}}
\end{figure*}

We start from the chiral N$^3$LO NN interaction of 
Refs.~\cite{Entem:2003th,Machleidt:2011bh}, with cutoff $\Lambda_{\rm NN}=500\,\MeV$ and 
apply a free-space SRG evolution to lower the momentum resolution scale, $\lambdaSRG$. 
The NN+3N-induced Hamiltonian consistently includes three-nucleon forces induced by the 
evolution. Results for this interaction correspond to the 
unevolved NN interaction, up to truncated induced 4N,\ldots,AN forces 
\cite{ppnpreview,Jurgenson:2009bs}. The NN+3N-full Hamiltonians also include an initial local 
chiral 3N interaction at  order N$^{2}$LO \cite{Navratil:2007ve}, consistently evolved to $\lambdaSRG$. We consider two cutoffs for the initial 3N interaction, $\Lambda_{\rm 3N} = 400, 500 \MeV$. The latter is naively consistent with $\Lambda_{\rm NN}$, although the NN interaction uses non-local regulators. Due to the reduced cutoff, the former avoids induced 4N interactions as $\lambdaSRG$ is lowered \cite{Roth:2012qf,Roth:2013kx}. The Hamiltonian has 27 low-energy constants in the NN interaction
plus two from the 3N interaction, which are fit to properties of few-body 
systems only, thereby providing predictions when applied to the medium-mass region.




The SRG-evolved
Hamiltonians are transformed to an angular-momentum-coupled basis built from single-particle 
HO states with quantum numbers $e=2n+l\leq\eMax$. An additional cut 
$e_{1}+e_{2}+e_{3}\leq \EMax < 3\eMax $ is introduced to manage storage of the 3N matrix 
elements. We use $\EMax=14$, which for resolution scales $\lambdaSRG=1.88-2.24\,\fm^{-1}$ 
contributes less than 1\% to the uncertainty of ground-state energies
\cite{Roth:2012qf,Binder:2013zr,Hergert:2013mi,Hergert:2013ij,Binder:2013uq}.


We first solve the HF equations for the $^{16}$O core using the intrinsic kinetic energy,
\begin{equation}
  \Tint = \TO-\TO_\text{cm}=\left(1-\frac{1}{A}\right)\sum_i\frac{\pOV_i^2}{2m}-\frac{1}{Am}\sum_{i<j}\pOV_i\cdot\pOV_j\,,
\end{equation}
with $A$ being the particle number of the \emph{target nucleus} to account for the change of 
the single-particle wavefunctions as $A_c \to A$ \cite{Hergert:2009wh}.  The Hamiltonian is 
then normal ordered with respect to the core's HF reference state, and the resulting in-medium 
zero-, one-, and two-body operators serve as the initial values for the IM-SRG flow equations. 
The residual three-body term is neglected, giving rise to the normal-ordered two-body (NO2B) 
approximation \cite{Hagen:2007zc,Roth:2012qf,Binder:2013zr}. 
The one- and two-body parts of the fully decoupled valence-space Hamiltonian are taken as the 
single-particle energies (SPEs) and two-body matrix elements to be diagonalized in a standard 
shell-model calculation, which we diagonalize in the $sd$ valence-space 
above an inert $^{16}$O core.


\begin{table}
\centering
\caption{IM-SRG $sd$-shell SPEs (in MeV) for $\lambdaSRG = 1.88 \fmi$ and
$\hw$ $=$ $24\MeV$, compared with MBPT \cite{Holt:2011fj} (NN+3N, see text) and phenomenological USDb values \cite{USD}. }
\label{spetab}
\begin{tabular*}{0.48\textwidth}{@{\extracolsep{\fill}}ccccccc}
\hline \hline
  Orbit        & NN & NN+3N-ind. & \multicolumn{2}{c}{NN+3N-full} & MBPT & USDb  \\ [.03in] 
                  &       &                     & $400\MeV$ & $500\MeV$ & & \\
  \hline
  $\rm d_{5/2}$ & $-7.07$ & $-3.77$ & $-4.62$ & $-7.14$ & $-3.78$ & $-3.93$ \\ 
  $\rm s_{1/2} $& $-5.80$ & $-2.46$ & $-2.96$ & $-4.42$ & $-2.42$ & $-3.21$ \\ 
  $\rm d_{3/2}$ & \,\,\, 1.81 & \,\,\, 2.33 & \,\,\, 3.17 & \,\,\, 2.85 & \,\,\, 1.45 & \,\,\, 2.11 \\ [.03in] \hline\hline
 \end{tabular*} 
\end{table}

Of interest is the $\hw$ dependence of the spectra, since $\hw$ is adjusted to the core in  
phenomenological shell-model calculations. We illustrate the effect of varying $\hw$ from 
$20\,\MeV$ to $24\,\MeV$ by shaded bands in all plots. Since this variation probes the 
convergence of the calculation and is mainly governed by $\lambdaSRG$ rather than the input 
Hamiltonian, we only show bands for the NN+3N-full Hamiltonians. Finally, coupling to the 
continuum is relevant in neutron-rich oxygen isotopes \cite{Volya:2005,Hagen:2012a}, and we 
will include these effects in future work. We indicate in all spectra the location of the neutron-
separation threshold to highlight the energy region where continuum is expected to become 
important.

The IM-SRG SPEs 
are given in Table \ref{spetab}. We compare to SPEs calculated in MBPT, from softened N${}^3$LO 
NN and re-fit N${}^2$LO 3N interactions \cite{Otsuka:2009,Holt:2010,ppnpreview}, and from
phenomenological USDb \cite{USD}.  The NN case, where the IM-SRG SPEs are 
significantly overbound, requires induced 3N to improve the description. For NN+3N-full, 
the $\Lambda_{\rm 3N}=400\MeV$ SPEs are comparable to MBPT and phenomenology, while 
those from $\Lambda_{\rm 3N}=500\MeV$ are more deeply bound.  The $\df-\dt$ gap is 
approximately $2\MeV$ ($4\MeV$ for $\Lambda_{\rm 3N}=500\MeV$) larger in IM-SRG, 
pointing to a stronger spin-orbit component than in MBPT.

Because of the deeply bound SPEs for $\Lambda_{\rm 3N}=500\MeV$, excited states lie 
$1.0-1.5\MeV$ higher in energy than for $\Lambda_{\rm 3N}=400\MeV$. This variation should 
be regarded only as a first step towards estimating the uncertainty, since it conflates 
uncertainties from the input Hamiltonian with those from the evolution to $\lambdaSRG$. For 
$\Lambda_{\rm 3N}=400\MeV$, these uncertainties are controlled, and we can directly 
confront the input chiral Hamiltonian with experiment 
\cite{Binder:2013zr,Hergert:2013ij,Hergert:2013mi,Roth:2013kx,Binder:2013uq}. For this
reason, we only discuss the spectra for this Hamiltonian in detail in the following.

The importance of 3N forces in determining the oxygen dripline was first highlighted in 
microscopic valence-space calculations \cite{Otsuka:2009,Holt:2011fj,Caesar}, then in 
\emph{ab initio} studies \cite{Hagen:2012a,Cipollone,Hergert:2013ij}. The evolution of the 
IM-SRG SPEs with the neutron number (Fig.~\ref{gs}(a)) reveals the same mechanism: in the 
NN+3N-induced case, the $\dt$ orbit remains bound past $^{20}$O. The repulsive effects of 3N 
forces shift the $\dt$ orbit to a higher starting point of $3.17\MeV$ in $^{17}$O, while 
moderating its decrease to neutron-rich isotopes, where it is bound by only $160\keV$ in 
$^{24}$O.  

We diagonalize the $A$-dependent IM-SRG valence-space Hamiltonian to obtain the
ground-state energies of $^{18-28}$O, shown in Fig.~\ref{gs}(b), which include the 
IM-SRG-calculated core energy. For an SRG-evolved NN interaction, the oxygen isotopes are 
overbound due to neglected initial and induced 3N forces, leading to unrealistic predictions. 
Including induced 3N forces lessens the overbinding, but fails to give the correct trend past 
$^{24}$O.  With initial 3N forces, agreement with experimental data is further improved, with 
moderate overbinding past $^{22}$O.  Moreover, the flat trend of the ground-state energies 
beyond $^{24}$O is similar to experimental data in $^{25,26}$O 
\cite{Kanungo,Lunderberg,Caesar} and agrees with other calculations based on chiral NN+3N 
forces \cite{Hagen:2012a, Cipollone,Otsuka:2009,Holt:2011fj}.  We note that $\nuc{O}{25-28}$ 
are weakly bound w.r.t. $\nuc{O}{24}$.  Finally, $\Lambda_{\rm 3N}=500\MeV$ ground-state 
energies are overbound with more pronounced $\lambdaSRG$ dependence. For instance in 
$^{22,24,28}$O, energies increase, respectively, to 
$-202.18\MeV$, $-215.42\MeV$, $-219.94\MeV$ for $\lambda_{\rm SRG}=1.88\fmi$ and 
$-193.31\MeV$, $-204.30\MeV$, $-206.85\MeV$ for $\lambda_{\rm SRG}=2.24\fmi$.

In contrast, the multi-reference IM-SRG (MR-IM-SRG) \cite{Hergert:2013ij} gives a robust 
prediction of the dripline at $\nuc{O}{24}$ for the NN+3N-full Hamiltonians, with ground-state 
energies in good agreement with experimental data and other \emph{ab initio} methods 
\cite{Hergert:2013ij}. The MR-IM-SRG evolution is carried out in the target nucleus rather than 
in the core with shifted $A$, so its open-shell reference state accounts for 
wavefunction-relaxation effects and the presence of nucleons in the valence space during the 
evolution.  Therefore differences like the observed $<2\%$ are to be expected. We will revisit 
the issue of over-binding by using $\nuc{O}{22}$ and $\nuc{O}{24}$ cores and compare these
results with MR-IM-SRG calculations of excited states.

\begin{figure}[t]
\includegraphics[scale=.33,clip=]{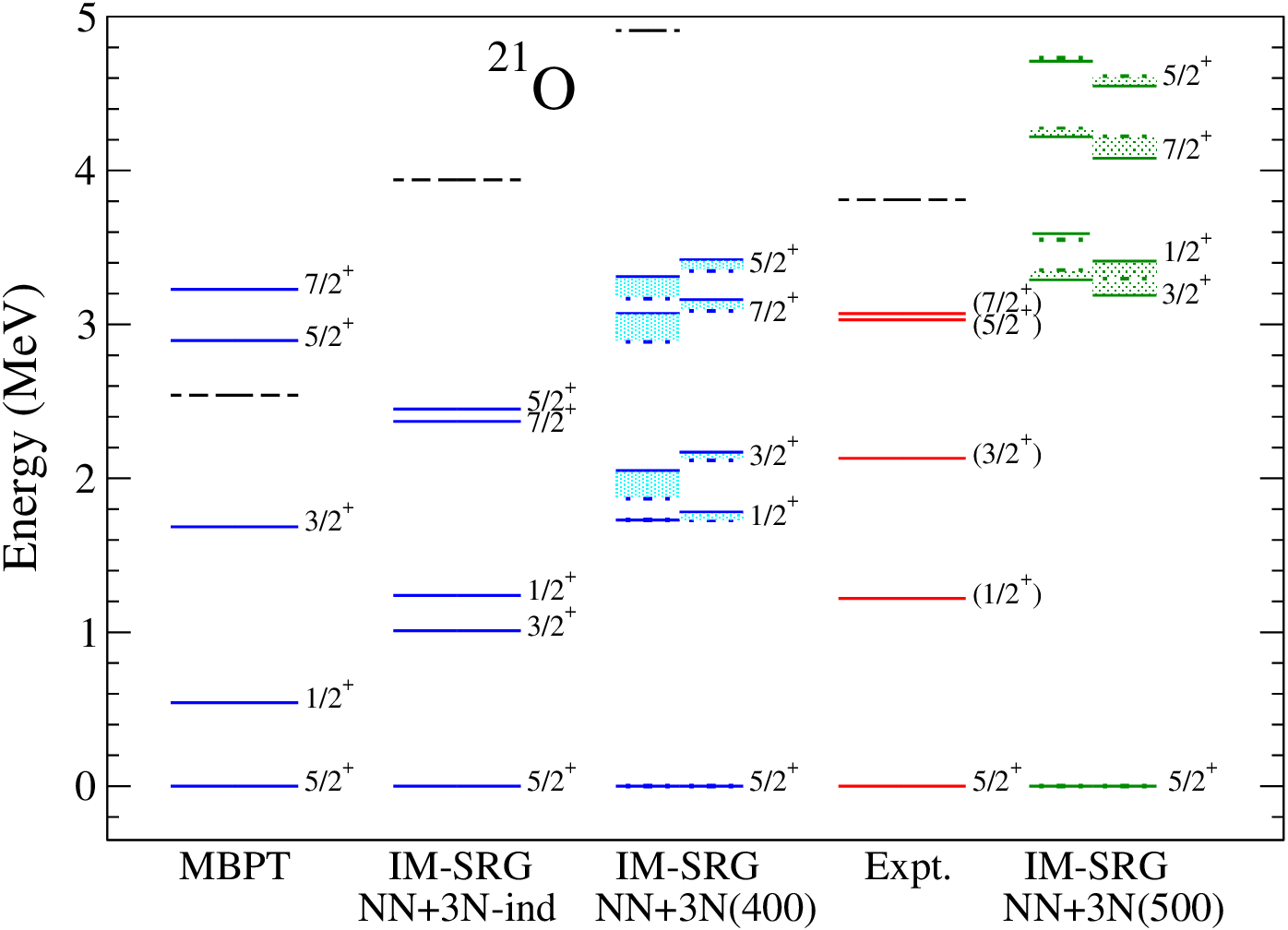}
\caption{Excited-state spectrum of $^{21}$O for IM-SRG Hamiltonians based on 
NN+3N-induced and NN+3N-full for $\Lambda_{\rm 3N}=400,500\MeV$, with $\hw=20\MeV$ (dotted) and $\hw=24\MeV$ 
(solid), compared with MBPT (NN+3N) and experiment \cite{Stanoiu}. NN+3N-full results are given in two columns: $\lambda_{\rm SRG}=1.88\fmi$ (left) and $\lambda_{\rm SRG}=2.24\fmi$ (right).  The neutron-separation 
threshold is given by black dot-dashed lines.  \label{21O}}
\end{figure}

In Fig.~\ref{gs}(b), we highlight the insensitivity of the ground-state energies to variation of 
$\hw$ from $20\,\MeV$ to $24\MeV$ in the band for the NN+3N-full Hamiltonian.  Differences 
only become non negligible for $A>24$. 
The weak dependence of calculated observables on $\hw$ is a 
striking feature of the nonperturbative IM-SRG valence-space approach, implying a 
remarkable level of convergence at the IM-SRG(2) level.

\begin{figure}
\includegraphics[scale=.33,clip=]{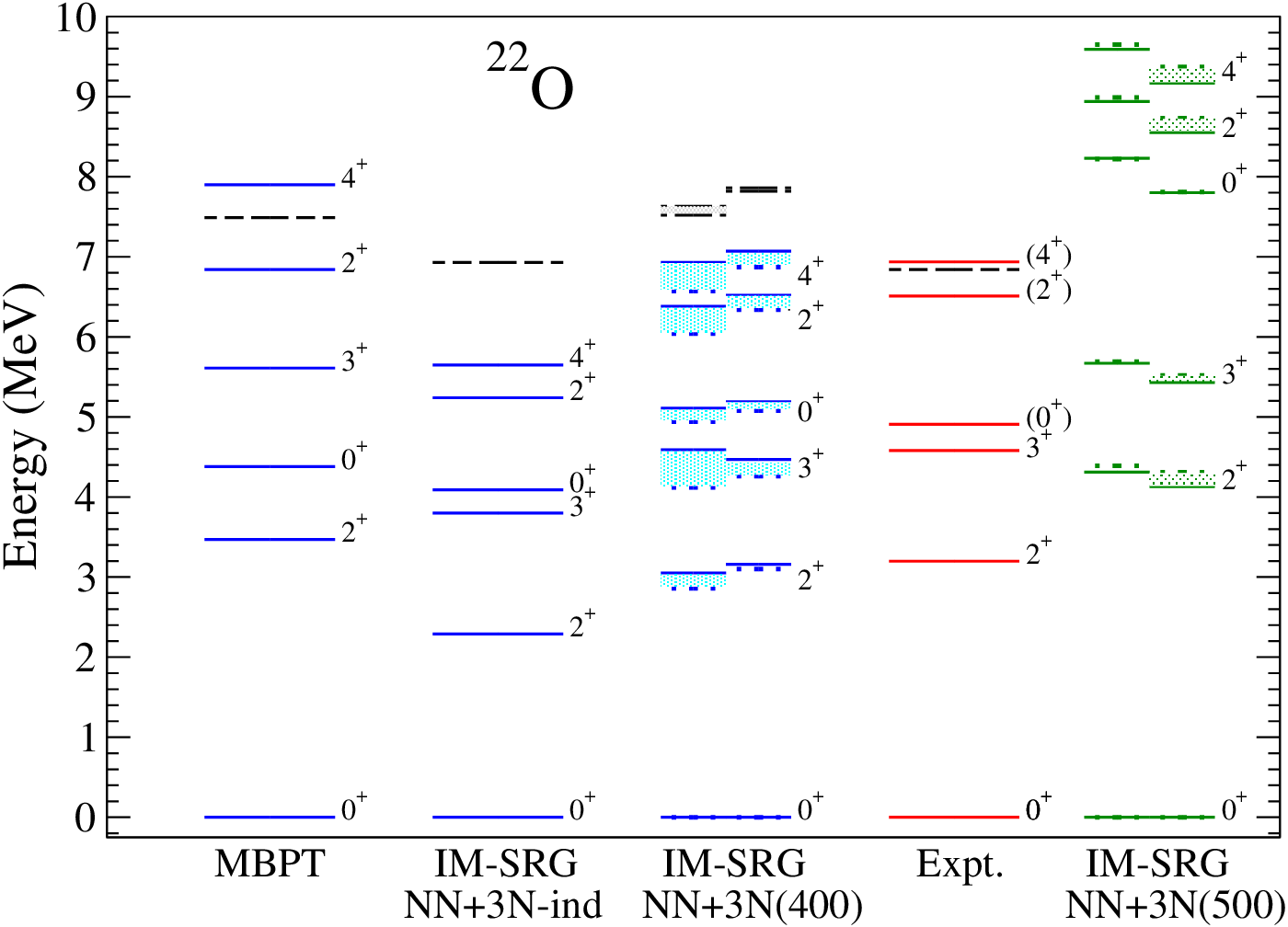}
\caption{Excited state spectrum for $^{22}$O as in Fig.~\ref{21O}, with experimental values from \cite{Stanoiu,Fernandez}.
\label{22O}}
\end{figure}


Turning to excitation spectra, since NN forces give a reasonable 
description of low-lying spectra near $^{16}$O \cite{MBPT}, we focus on the region of the new 
$N=14,16$ magic numbers, and the limit of stability by considering $^{21-26}$O. As shown in 
Ref.~\cite{Holt:2011fj}, microscopic valence-space Hamiltonians from MBPT, calculated in the 
standard $sd$-shell, do not adequately reproduce experimental data, even with 3N forces.  
With the inclusion of the $\fs$ and $\pt$ orbitals, spectroscopy improves, indicating that a 
perturbative treatment of these orbitals may be insufficient.  Given the nonperturbative
character of IM-SRG, we expect similar improvements already in the $sd$ shell.  

We first consider the spectrum of $^{21}$O in Fig.~\ref{21O}.  While no calculation fully 
reproduces experiment, MBPT and IM-SRG correctly predict the ordering of the first two excited 
states with an initial 3N force, but the $1/2^+$ level lies too low in MBPT and too high in 
IM-SRG. Since the MBPT results are obtained with a softened N${}^3$LO interaction with a 
re-fit 3N interaction \cite{Otsuka:2009,Holt:2010}, disagreements are both due to the different 
input Hamiltonians and the nonperturbative IM-SRG approach. We note that the tentative 
$7/2^+$ and $5/2^+$ assignments are reproduced in both calculations, but the ordering is 
reversed in IM-SRG compared to MBPT. 

The calculated IM-SRG spectra of $^{22}$O are compared with MBPT \cite{Holt:2011fj} and 
experiment \cite{Stanoiu,Fernandez} in Fig.~\ref{22O}.  Without initial 3N forces, the spectrum 
is too compressed. The $\tpi$ state, in particular, is $1.0\MeV$ too low, contradicting the doubly 
magic nature of $^{22}$O.  Unlike MBPT or the phenomenological USDb Hamiltonian, the 
IM-SRG reproduces the correct ordering of the $3^+_1$ and $0^+_2$ states. Inclusion of 
initial 3N forces leads to significant improvement, and the final spectrum is very close to 
experiment. The extended-space MBPT calculations reproduce the high $\tpi$ state 
but have too uniform spacing and incorrect $3^+_1-0^+_2$ ordering.

\begin{figure}[t]
\includegraphics[scale=.33,clip=]{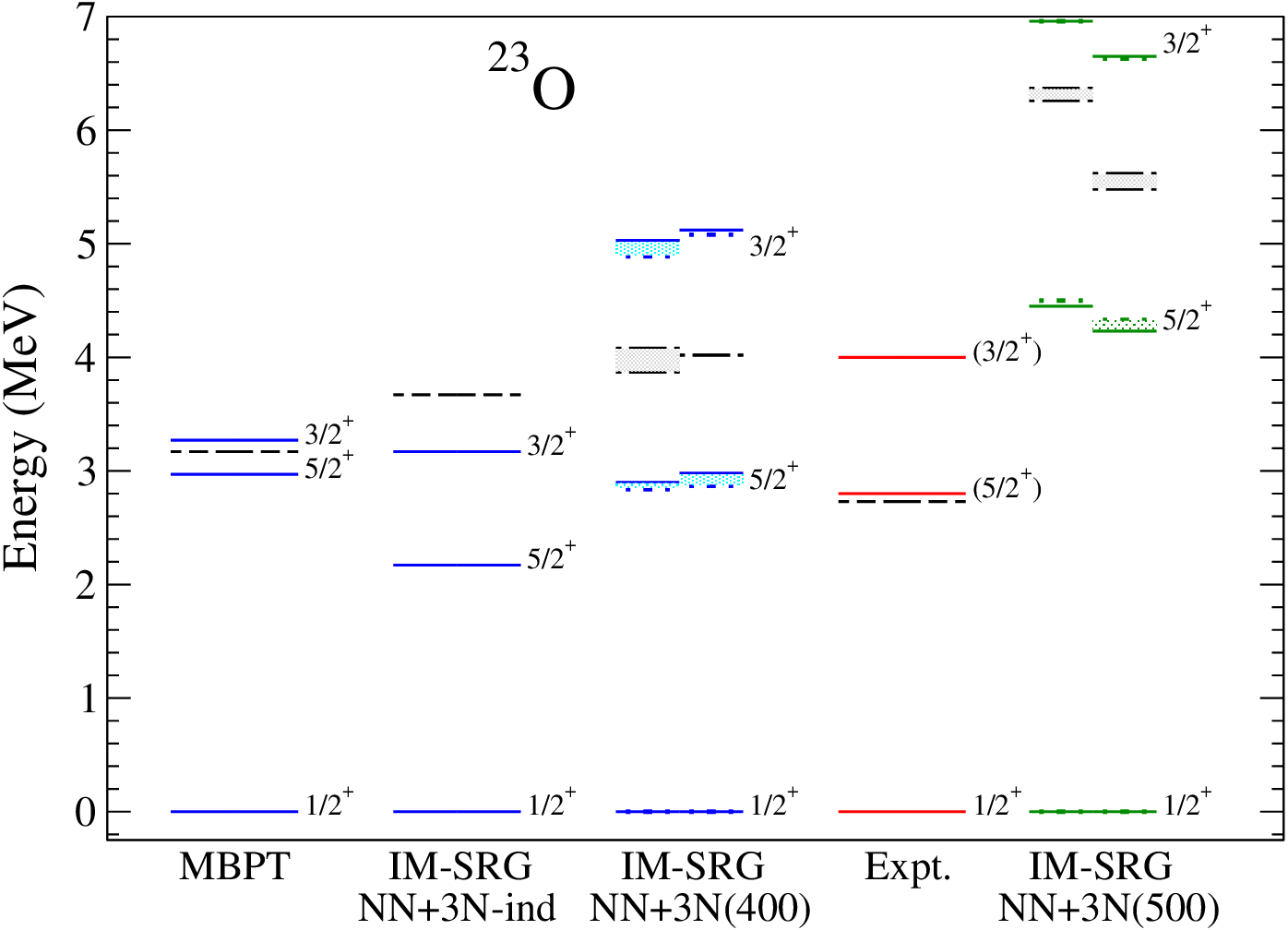}
\caption{Excited state spectrum of $^{23}$O as in Fig.~\ref{21O}, with experimental values 
from \cite{Elekes, Schiller}.
\label{23O}}
\end{figure}

\begin{figure}[t]
\includegraphics[scale=.33,clip=]{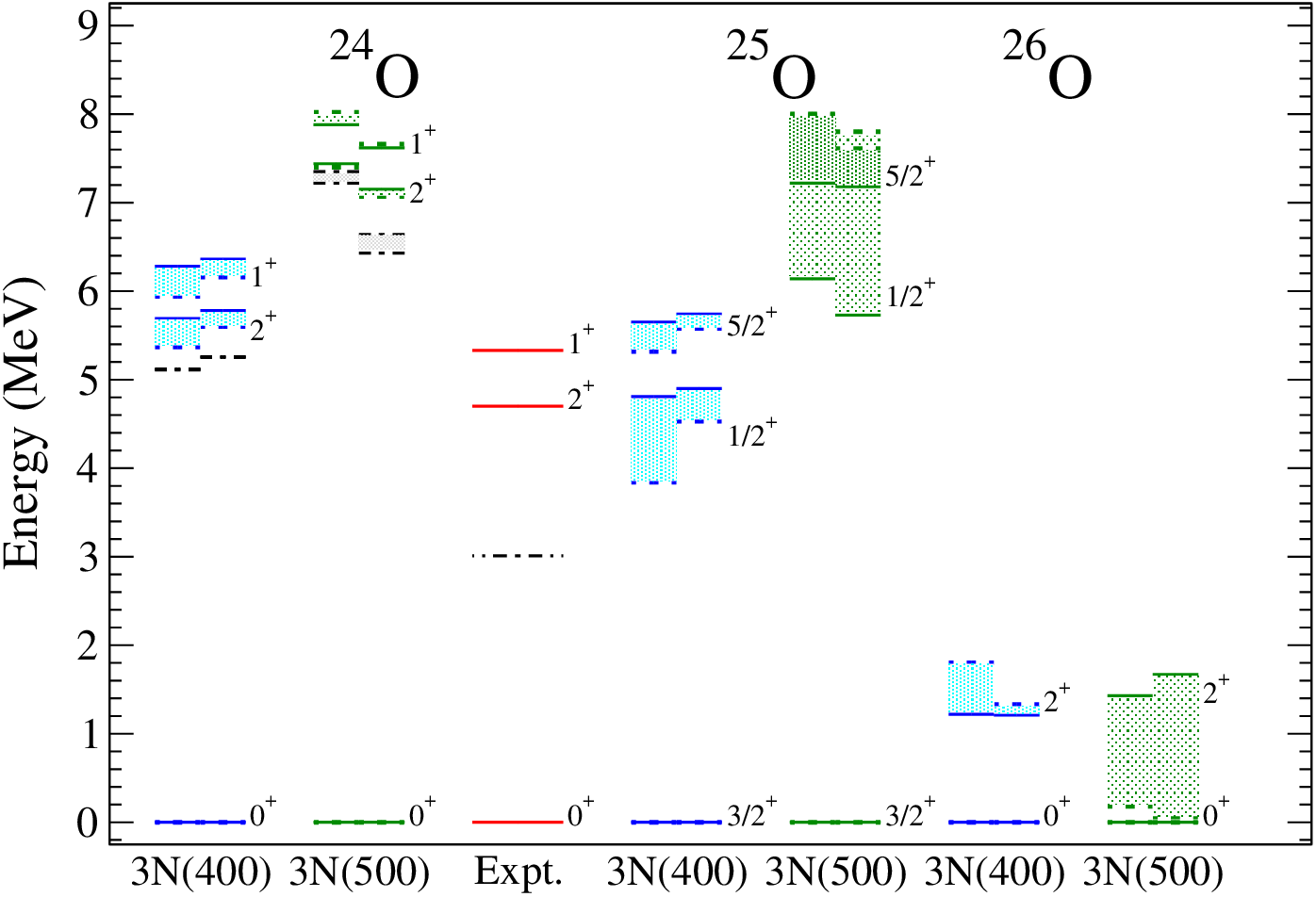}
\caption{Excited state spectra of $^{24-26}$O for $\Lambda_{\rm 3N}=400,500\MeV$ NN+3N-full Hamiltonians with $\hw=20\MeV$ (dotted), $\hw=24\MeV$ (solid), and $\lambda_{\rm SRG}$ dependence as in Fig.~\ref{21O}, compared to experiment for $^{24}$O \cite{Hoffman:2009,Tshoo:2012}.
\label{24O}}
\end{figure}

There are no bound excited states in $^{23}$O, only two higher-lying states, tentatively 
identified as $5/2^+$ and $3/2^+$, indicating the sizes of the $\df-\so$ and $\dt-\so$ gaps,
respectively \cite{Elekes,Schiller}. We show the calculated and experimental spectra in 
Fig.~\ref{23O}. Again, the IM-SRG does not reproduce this spectrum without initial 3N forces: 
the $5/2^+$ state is nearly $1\MeV$ too low, but the $5/2^+-3/2^+$ gap close to experiment.  
Similar to MBPT with initial 3N forces, the $5/2^+$ energy is almost exactly that of experiment, 
only the $3/2^+$ is $1\MeV$ too high. Due to its position $2\MeV$ above threshold, it is 
expected that continuum effects will lower this state, bringing it closer to the experimental value.

As expected from the high $3/2^+$ state in $\nuc{O}{23}$, we also predict $^{24}$O to be 
doubly magic, but with a $\tpi$ energy $1.2\MeV$ higher than experiment, as seen in 
Fig.~\ref{24O}.  Nonetheless, the $2^+-1^+$ spacing is very close to experiment, and with 
continuum effects included, these states will be lowered. Finally, we present predictions for 
excited-state energies in the unbound $^{25,26}$O isotopes.  We again find large $1/2^+$ and 
$5/2^+$ excitation energies in $^{25}$O, which are expected to decrease with continuum 
coupling. In $\nuc{O}{26}$, we predict one low-lying state below $6\MeV$: a $2^+$ near 
$2\MeV$. A tentative identification of an excited state near $4\MeV$ was reported in 
\cite{Caesar}, but no such natural-parity state was found in our calculations. 


We have presented the first \emph{ab initio} construction of a nonperturbative $sd$-shell Hamiltonian 
based on chiral NN and 3N forces.  The SPEs and two-body matrix elements are well converged with 
respect to basis size and exhibit weak $\hw$ dependence.  Furthermore, a good description of ground 
and excited states is found throughout the oxygen isotopes in a valence space consisting of only the 
$sd$-shell orbits.  This provides the exciting possibility to extend these calculations to nearby F, Ne, and 
Mg isotopic chains and through extending the valence space, will give access to the island-of-inversion 
region and potentially the full $sd$-shell neutron dripline.  

\paragraph{Acknowledgments.}
We thank R. Furnstahl, M. Hjorth-Jensen, C. R. Hoffman, and J. Men\'endez for useful discussions.
This work was supported in part by the NUCLEI SciDAC Collaboration under the U.S. Department of 
Energy Grants No. DE-SC0008533 and DE-SC0008511, the National Science Foundation under Grants 
No.~PHY-1002478, PHY-1306250, and PHY-1068648, the Helmholtz Alliance Program of the Helmholtz 
Association, contract HA216/EMMI ``Extremes of Density and Temperature: Cosmic Matter in the 
Laboratory'', the DFG through grant SFB 634, the BMBF under Contract No. 06DA70471 the ERC (grant 
307986 STRONGINT), and HIC for FAIR. Computing resources were provided by the Ohio Supercomputing Center (OSC) and by the J\"ulich Supercomputing Center (JUROPA).

%
%
\bibliography{2013_imsrg_sm}

\end{document}